\documentstyle[11pt,amssymb]{article}

\def\be{\begin{equation}}
\def\ee{\end{equation}}

\title{{\hfill\small\tt Funct.\,Anal.\,Appl.\,{\bf 12},\,121-128\,(1978)}\\
{\hfill}\\
{Quantum Systems Related to Root Systems,}\\ 
{and Radial Parts of Laplace Operators}}
\author{M.A. Olshanetsky and A.M. Perelomov\\
{\small\em Institute for Theoretical and
Experimental Physics, 117259 Moscow, USSR}}
 
\date{}
\begin{document}
\maketitle{}

%\begin{abstract}\noindent
%\end{abstract}

\section{Introduction}
\setcounter{equation}{0} 

Laplace operators on semisimple Lie groups and symmetric spaces 
were studied by Gel'fand [1,2]. The explicit form of the radial parts of 
Laplace operators (RPLO) was calculated only for symmetric spaces with a 
complex group of motions [3] and for Cartan domains of the first type [4]. 

In this paper we shall present the explicit form of generating algebras 
of RPLO for symmetric spaces that have a bounded system of roots of type 
$A_n$, and for all symmetric spaces of rank 2, with the exception of 
$G_2\,/\,(SU(2) \otimes SU(2))$. These results can be obtained by studying a 
quantum system constructed on the basis of root systems by a method 
proposed by the authors. The corresponding classical systems were studied 
previously [5].

The connection between dynamical systems and symmetric spaces is useful in 
studying both of them.

In particular, it can be proved that for certain values of the constants 
occuring in the Hamiltonian, the quantum systems and the classical systems 
are completely integrable for all the systems of roots.

Let us note that by constructing completely integrable quantum systems, we 
obtain an explicit description of a commutative algebra of differential 
operators of several variables.

The authors express their gratitude to F.I. Karpelevich for a very useful 
discussion of this paper.

\section{Quantum Systems Related to Root Systems}
\setcounter{equation}{0}

Let $R$ be a root system in the space ${\cal H}$ (dim\,${\cal H}=n$), and let 
$R_+$ be a subsystem of roots that are positive under a certain order 
(the definitions and terms used here can be found in [6]).

Let $q\in {\cal H}$ and let $q_\alpha =(q,\alpha )$ be the scalar product 
of $q$ and of the root $\alpha $. As usual, we shall introduce the operators 
of the coordinate $q_j$ and of the momentum $p_j=-\,i\,\partial /
\partial q_j$, and let $g_\alpha ^2$ be positive constants on roots of equal 
length. As in [5], we shall consider systems with a Hamiltonian of the form 
\be H=\frac{p^2}2+U(q), \ee
where
\be U(q)=\sum _{\alpha \in R_+} g_\alpha ^2\,V(q_\alpha ), \ee
and the function $V(\xi )$ can have one of the following forms:
\begin{eqnarray}
V(\xi ) &=& \nonumber \\
\mbox{I.} && \xi ^{-2},\nonumber \\
\mbox{II.} && a^2\,\mbox{sh}^{-2}a\xi ,\nonumber \\
\mbox{III.} && a^2\,\mbox{sin}^{-2}a\xi ,\\
\mbox{IV.} && a^2\,\gamma (a\xi ), \nonumber \\
\mbox{V.} && \xi ^{-2}+\omega ^2\xi ^2. \nonumber \end{eqnarray}
For a root system of type $A_{n-1}$ we have $q_\alpha =q_k-q_l$ ($k\neq l$). 
In this case,
\be U(q)=g^2 \sum _{k<l}V(q_k-q_l). \ee

Such quantum systems describe $n$ particles on a straight line; they were 
studied in [7-13]. A quantum system of type $G_2$ has been considered in 
[14], [15].

Let us note that although the Hamiltonian (1) is defined in the space 
${\cal H}$, it is in fact impossible for a particle to move in the entire 
space ${\cal H}$.

Let $\Lambda $ be a Weyl chamber defined on the basis of a subsystem of 
positive roots,
\be \Lambda =\{ q\in {\cal H}|q_\alpha =(q,\alpha )>0,\,\alpha \in R_+\}, \ee
and $\Lambda _a$ be the Weyl alcove
\be \Lambda _a =\{ q\in {\cal H}|0<aq_\alpha <\pi ,\,\alpha \in R_+\}.\ee

Then the potential $U(q)$ in (2) becomes infinite on the walls of the Weyl 
chamber $\Lambda $ (5) for systems of type I, II and V, and on the walls 
of the alcove $\Lambda _a$ (6) for systems of type III. It is possible 
to indicate  an alcove form also for the functions $a^2\,\gamma (a\xi )$ 
(IV (3), see [5]). Thus we have found that for the systems I, II and V 
the particle moves in the Weyl chamber $\Lambda $, whereas for the systems 
III and IV it moves in the Weyl alcove $\Lambda _a$.

Now let us present two definitions.

{\bf Definition 1}. {\em A differential operator that commutes with the 
Hamiltonian} $H$ {\em is called an integral of motion} (IM).

It is evident that the IM form an algebra.

{\bf Definition 2}. {\em A quantum system with a Hamiltonian} $H$ {\em 
is said to be completely integrable if the algebra} IM {\em contains 
a commuataive subalgebra that has at least} $n$ {\em functionally 
independent operators} ($n$ = dim\,${\cal H}$).

\section{The Mapping $j$}
\setcounter{equation}{6}

In this section we shall study a mapping $j$ in the algebra of differential 
operators that carries the Laplace--Beltrami operator into the Hamiltonian 
operator of a quantum system.

Let us note that this mapping has been used in [3] and [4] for the 
calculation of the RPLO.

At first we shall consider symmetric spaces of negative curvature $X^-$. 
If $G$ is a semisimple Lie group with a finite center, and $K$ is its 
maximal compact subgroup, then $X^-=G/K$. The Laplace operators are 
$G$-invariant differential operators on $X^-$ [16].

Let ${\cal G}$ and ${\cal K}$ be Lie algebras of the groups $G$ and $K$, and 
let ${\cal L}$ be an orthogonal complement of ${\cal K}$ in ${\cal G}$ in 
the sense of Cartan's scalar product. By ${\cal H}$ we shall denote the 
Cartan subalgebra in ${\cal L}$. (The subalgebra ${\cal H}$ is the vector 
space introduced in Sec. 2).

Let $R=\{\alpha \}$ be a bounded system of roots of the space $X^-$ [18]. 
The roots $\alpha \in R$ are linear forms on ${\cal H}$. If $\Lambda $ 
is a fixed Weyl chamber of the root system $R$, and $x_0k=x_0$ for $k\in K$, 
then for any point $x\in X^-$ there exists an unique element $q(x)\in 
\Lambda $ such that
\be x=x_0\,\exp \{q(x)\}\,k,\qquad k\in K \ee
(the Cartan decomposition).

Functions that are invariant under rotations ($f(xk)=f(x)$) depend only on 
$q(x)$. On these functions the Laplace operators induce their radial part. 
Let us denote the RPLO algebra by $D$. The RPLO is $W$-invariant ($W$ being 
Weyl's group).

Let $m_\alpha $ be the multiplicity of the root $\alpha $, let $R_+$ be a 
subsystem of roots in $R$ that is positive with respect to $\Lambda $, 
and let $x(q_\alpha )=a\,\mbox{sh}^{-1}aq_\alpha $ ($V(q_\alpha )=x^2
(q_\alpha )$)\,\footnote{\,\,For simplicity, in intermediate calculations 
the parameter $a$ will be set equal to unity.}.

On $\Lambda $ let us define a function $\xi (q)$:
\be \xi (q)=\prod _{\alpha \in R_+} x(q_\alpha )^{-m_\alpha /2}\,. \ee
Let us consider a mapping $j$ of the algebra $T$ of all differential 
operators on ${\cal H}$:
\be j\colon t \to \xi (q)\,t\xi ^{-1}(q),\qquad t\in T. \ee
Let us define the operator $B\in D$ as 
\be B=-\,\xi ^{-2}(q)\,\sum _{l=1}^n {\hat p}_l\,\xi ^2(q)\,{\hat p}_l. \ee
This oparator is the radial part of the Laplace--Beltrami operator on the 
space $X^-$.

{\bf Lemma 1.}\,\footnote{\,\,In the particular case $m_\alpha =1$ and root 
systems of type $A_n$, we can find in [17] a remark that corresponds to this 
lemma.} {\em If $H$ is the Hamiltonian} (1) {\em with a function} $V(q)$ 
{\em of type} II (3), {\em then}
\be H=-\,j\left[ \frac12\left(B+\rho ^2\right)\right],\qquad \rho =\frac12\,
\sum _{\alpha \in R_+} m_\alpha \alpha , \ee
{\em the constants} $g_\alpha $ {\em in} (2) {\em being related to the root 
multiplicities} $m_\alpha $ {\em as follows}:
\be g_\alpha ^2=\frac18\,m_\alpha \,(m_\alpha +2\,m_{2\alpha }-2)\,
|\alpha |^2. \ee

{\bf Proof.}  According to (10) and (11), we have to show that
\be H=-\,\frac12\,\xi (q)\left( B+\rho ^2\right) \xi ^{-1}(q).\ee
Let us calculate the operator in the right-hand side. According to (10), 
 we obtain
\be -\,\frac12\,\xi (q)\left( B+\rho ^2\right) \xi ^{-1}(q)=\frac12\,
\xi ^{-1}(q) \sum _{l=1}^n p_l\xi ^2(q)\,p_l\xi ^{-1}(q)-\frac{\rho ^2}2
=\frac12\,p^2+U(q)\,, \ee
where
\be U(q)=\frac12 \left( \sum _{j=1}^n \xi ^{-1}(q)\,\xi _{jj}(q)-\rho ^2
\right) \ee
and $\xi _{jj}$ is the second derivative with respect to $q_{j}$. It can be 
asserted  that this expression coincides with the potential (2). 
By substituting into (15) the explicit form of the function $\xi (q)$ (8), 
we obtain
\be  U(q)=\sum _{\alpha \in R_+} \frac{m_\alpha }{4}\left( \frac{m_\alpha }2
-1\right) |\alpha |^2\,x^2(q_\alpha ) + F(q), \ee
where
\be F(q)=\frac18\,\sum _{\alpha ,\beta \in R_+,\,\alpha \neq \beta }
m_\alpha m_\beta (\alpha, \beta )\,[\mbox{cth}\,q_\alpha \,\mbox{cth}\,
q_\beta -1]. \ee

Let $R$ be a reduced system of roots. This signifies that $m_\alpha 
m_{2\alpha }=0$. Let us show that in this case,
\be F(q)\equiv 0. \ee

If (18) holds, then the assertion of the lemma follows from (16) and (13), 
(14).

From (7) it follows that (18) is equivalent to the vanishing of the function 
$\Phi (q)=F(q)\,\xi ^{(0)}(q)$, where $\xi ^{(0)}(q)=\prod _{\alpha \in R_+} 
\mbox{sh}\,q_\alpha $. The function $\Phi (q)$ has the following form:
\be \Phi (q)=c\sum _{\alpha ,\beta \in R_+,\,\alpha \neq \beta } m_\alpha 
m_\beta (\alpha ,\beta )\left( e^{q_\alpha -q_\beta }+e^{q_\beta -q_\alpha }
\right) \prod _{\gamma \in R_+,\,\gamma \neq \alpha ,\,\gamma \neq \beta }
\left( e^{q_\gamma }-e^{-q_\gamma }\right) . \ee

Let $W$ be Weyl's group of the root system $R$. Let us note that the 
function $F(q)$ is $W$-invariant by virtue of its construction (see (16)), 
whereas the function $\xi ^{(0)}(q)$ is antiinvariant. Hence, $\Phi (q)$ is 
also antiinvariant.

Let us consider the group algebra $A(P)$ over the ring $R$ of the additive 
weight group $P$ of the root system $R$ (see [6], Chap.VI, Sec.3). Its 
elements are linear combinations of formal exponents $e^p$ ($p\in P$), and 
$e^pe^{p'}=e^{p+p'}$.

In particular, the element
\be \varphi =\sum _{\alpha ,\beta  \in R_+,\,\alpha \neq \beta } 
m_\alpha m_\beta (\alpha ,\beta )\left( e^{\alpha -\beta }+e^{\beta -\alpha }
\right) \prod_{\gamma \in R_+,\,\gamma \neq \alpha ,\,\gamma \neq \beta }
\left( e^\gamma -e^{-\gamma }\right) \ee
belongs to $A(P)$. Since $\Phi (q)=\varphi (q)$ (see (19)), it follows that 
$\varphi $ is antiinvariant. From (20) it follows that the maximal term of 
the element $\varphi $ (see [6], Chap.VI, Sec.3.2) is $e^{\rho -\alpha }$, 
where $\alpha \in R_+$. An antiinvariant element with such a maximal term 
vanishes identically. Therefore, the function $\Phi (q)$, and hence also 
$F(q)=\Phi (q)\,\xi ^{(0)}(q)$ is equal to zero.

Now let $R$ be an unreduced system of roots. There exists only one such 
system $BC_n$. This root system has subsystems of type $B_n$ and $C_n$. Let 
us denote by $\bar R$ a root subsystem of $R$ of the form ${\bar R}=\{ 
\alpha \in R_+|2\alpha \in R\}$. Then the function $F(q)$ (see (17)) can be 
expressed in the form
\be F(q)=F_{B_n}(q)+F_{C_n}(q)+\frac14\,\sum _{\alpha ,\beta \in {\bar R}} 
(m_\alpha \alpha ,m_{2\beta }\beta )(\mbox{cth}\,q_\alpha \,\mbox{cth}\,2q_
\beta -1), \ee
where $F_{B_n}$ ($F_{C_n})$ is a function $F(q)$ that corresponds to the root 
system $B_n$ ($C_n$). Since $B_n$ and $C_n$ are reduced root systems, it 
follows that $F_{B_n}=F_{C_n}=0$. Let us also note that if $\alpha ,\beta \in 
{\bar R}$, then $(\alpha ,\beta )=\delta _{\alpha \beta }$. Therefore 
we can rewrite (21) in the form
\[ 
F(q)=\frac12\,\sum _{\alpha \in {\bar R}} m_\alpha m_{2\alpha }(\mbox{cth}\,
q_\alpha \,\mbox{cth}\,2q_\alpha -1). \]
By substituting this expression into (16) and performing very simple 
trigonometric transformations, we obtain the required equation
\be U(q)=\frac12\,\sum _{\alpha \in R_+}\frac{m_\alpha }2\left( \frac
{m_\alpha }2+m_{2\alpha }-1\right) |\alpha |^2\,V(q_\alpha ). \ee

From this lemma and from the form of $j$ we obtain the following

{\bf Proposition 1.}
\begin{eqnarray}
1. &\qquad\qquad \mbox{Ker}\,j=0, \\
2. &\qquad\qquad j\quad \mbox{preserves}\,\,\mbox{leading}\,\,\mbox{terms}, \\
3. &\qquad\qquad j\quad \mbox{preserves}\quad W-\mbox{invariance}, \\
4. &\qquad\qquad j(D)\subset S\quad (\mbox{an}\,\,\mbox{algebra}\,\,
\mbox{of}\,\,\mbox{IM}). \end{eqnarray}

It is evident that (26) makes sense only for group values of $g_\alpha $ 
(12). The root multiplicities $m_\alpha $ that determine $g_\alpha $ can be 
found in [18] for all symmetric spaces.

{\bf Remark}. The analysis of this section is completely applicable also to 
systems of type I and III (3). For this purpose let us introduce the 
parameter $a$ into the function $\xi (q)$ (8)
\[ \xi _a(q)=\prod _{\alpha \in R_+}\left( a^{-1}\,\mbox{sh}\,aq_\alpha 
\right) ^{\mu _\alpha },\qquad \mu _\alpha =\frac12\,m_\alpha . \]
Let $a$ tends to zero. We obtain a mapping of the RPLO on symmetric 
spaces of zero curvature $X^0$ in an IM algebra of systems of type I. But 
if $a$ is imaginary, then we obtain systems of type III and spaces of 
positive curvature $X^+$ that are dual in the sense of Cartan to $X^-$. Let 
us note that the curvature of a symmetric space is proportional to $a^2$. 

{\bf Theorem 1.} {\em Quantum systems of type} I, II, III, {\em and} V 
{\em are completely integrable for group values of} $g_\alpha $.

{\bf Proof}. Since the algebra $\tilde S$=Im\,$j(D)$ is isomorphic to the 
algebra $D$ (23) and $\tilde S\subset S$ (26), the assertion of the 
theorem for the systems I, II and III will follow from the existence of $n$ 
generators of the commutative algebra $D$ on the symmetric spaces $X^0$, 
$X^-$, and $X^+$ [16]. For systems of type V the assertion follows from the 
complete integrability of systems of type I when we change the 
variables $p_j\to p_j\pm i\,\omega q_j$ (see [19]).

{\bf Corollary 1.} {\em For group values of} $g_\alpha $, {\em the classical 
systems of type} I, II, III, {\em and} V {\em are completely integrable.}

{\bf Proof.} Let us introduce the Planck constant into the momentum operators 
$p_j=-\,i\hbar\,\partial/\partial q_j$, and let it tends to zero. Then the 
commutators of the operators will go over into the Poisson brackets of 
classical variables. Let us note that in the classical case, complete 
integrability has been proved earler (in [5]) for classical root systems 
only.

\section{Algebra of Integrals of Motion}
\setcounter{equation}{26}

In this section we shall describe a subalgebra $\tilde S$ of the IM algebra 
$S$ which for systems of type I - III and group values of $g_\alpha $ (12) 
is an image of the RPLO algebra $D$ under the mapping $j$.

Let us define the subalgebra $\tilde S$ in a more general situation. Let us 
consider systems of type I - IV for any values of $g_\alpha $. The 
subalgebra $\tilde S$ is specified by the following conditions:
\begin{description}
\item[1.] \be I(sp,sq)=I(p,q)\qquad (W-\mbox{invariance}), \ee
\item[2.] For systems of type I, the algebra ${\tilde S}^I$ is graded,
\be {\tilde S}^I=\oplus _{k=0}^{\infty }{\tilde S}_k^I,\qquad {\tilde S}_k^I=
\{ I_k^I(p,q)\},\qquad 
I_k^I\left( \lambda ^{-1}p, \lambda q\right) =\lambda ^{-k}\,I_k^I(p,q), \ee
where $I_k^I(p,q)$ is a polynomial of degree not higher than $k$ in $p$.
\item[3.] For systems of type II - IV,
\be I(p,q)=I^I(p,q)\,(1+O(|q|))\qquad \mbox{for}\quad q\to 0, \ee
where $I^I(p,q)$ is an IM of systems of type I. This relation makes it 
possible to introduce the gradation in $\tilde S$ for systems II - IV.
\end{description}

Let $I_k(p,q)\in {\tilde S}_k$ and suppose that it has the form
\be I_k(p,q)=\sum _{m=0}^s c^{i_1,\ldots ,i_m}(q)\,p_{i_1}\cdots  p_{i_m}
\qquad (s\leq k). \ee

{\bf Lemma 2}. {\em If the leading coefficients} $c^{i_1,\ldots ,i_s}(q)$ 
{\em in} (30) {\em are not constant, then} $I_k(p,q)=0$.

{\bf Proof}. Since $I_k\in \tilde S_k$, it follows that $[H,I_k]=0$. Let 
the coefficients of the leading powers $p_{i_1}\cdots p_{i_{s+1}}$ of 
this commutator are equal to zero. By virtue of the explicit form of $H$ 
(see (1) and (2)), we obtain
\be \sum _\sigma \frac\partial {\partial q_l}\,c^{j_1,\ldots ,j_s}(q)=0, \ee
where the sum is taken over all the permutations of indices $\sigma (l,
j_1,\ldots ,j_s)=(i_1,\ldots ,i_{s+1})$. It is proved ([3], Lemma 2.5, 
p.407) that the system (31) has only polynomial solutions. For systems 
of type I, from the gradation conditions (28) it follows that 
$c^{i_1,\ldots ,i_s}(\lambda q)=\lambda ^{s-k}c^{i_1,\ldots ,i_s}(q)$. 
Since $k>s$, the only polynomial to satisfy this condition is the polynomial 
being equal to zero identically. Thus, $c^{i_1,\ldots ,i_s}(q)\equiv 0$. 
The same result can be obtained by considering the asymptotic behavior, 
for $|q|\to 0$ of systems of type II - IV in (29).

{\bf Corollary 2.} {\em The integrals} $I_k(p,q)\in {\tilde S}_k$ 
{\em are uniquely determined by their leading terms.}

{\bf Proof}. Suppose that $I_k'$ and $I_k''$ belong to ${\tilde S}_k$, and 
that their leading terms coincide. Then $\Delta I_k=I_k'-I_k''$ will belong 
to $S_k$ and it does not contain constant leading terms. Hence, $\Delta 
I_k=0$.

{\bf Corollary 3}. {\em The algebra} $\tilde S$ {\em is commuataive.}

{\bf Proof}. Let $I_k\in {\tilde S}_k$ and $I_m\in {\tilde S}_m$. It is 
evident that $[I_k,I_m]$ belongs to ${\tilde S}_{k+m}$ and that it does not 
contain constant leading terms. Therefore, the commutator will be equal to 
zero.

Now let us confine ourselves to systems of type I - III and to group values 
of the constants $g_\alpha $ (12).

{\bf Theorem 2.} {\em The mapping} $j$ {\em is an isomorphism of an} RPLO 
{\em algebra on symmetric spaces of zero, negative, and positive curvature 
into the algebra} ${\tilde S}$ {\em of systems of type} I, II {\em and} 
III, {\em respectively.}

{\bf Proof}. By virtue of Proposition 1, the mapping $j$ is a 
monomorphism of the algebra $D$ into an algebra of $W$-invariant IM.

The Lie algebra ${\cal G}$ of the group of motions of a space of zero 
curvature has a decomposition ${\cal G}={\cal K}+{\cal L}$, where ${\cal K}$ 
is a maximal compact subalgebra and ${\cal L}$ is a subalgebra of 
parallel translations of the space $X^0$. The Lie operators belonging to 
${\cal K}$ have degree 0 under similarity transformations, 
whereas the Lie operators belonging to ${\cal L}$ have degree 1. Therefore, 
an universal enveloping algebra has a natural gradation.

This gradation can be transferred to the algebra of Laplace operators on 
the space $X^0$. This follows from the fact that in canonical coordinates 
the Laplace operators have constant coefficients, and also from the 
homogeneity of Lie operators. Therefore, any homogeneous component of the 
Laplace operator is also  Laplace operator. This gradation can evidently 
be transferred to the RPLO algebra $D$. From the explicit form of $j$ 
(see (9)) it follows that $j$ preserves  this gradation.

Now let us consider symmetric spaces of nonzero curvature.

Let the point $x$ tends to $x_0$ (see (7)). This corresponds to $q\to 0$. 
The space which is tangent at the point $x_0$ to $X$ is a symmetric space 
of zero curvature $X^0$. Hence, for $x\to x_0$ the Lie operators on $X$ 
will go over into Lie operators on $X^0$. It hence follows that the RPLO 
on $X$ has an asymptotic behavior (29). Let us note that the mapping $j$ 
preserves this asymptotic behavior. Thus, $j$ is a monomorphism of $D$ into 
${\tilde S}$.

Let $I_k\in {\tilde S}_k$. Let us consider the operator $j^{-1}\,I_k$. By 
virtue of Proposition 1 we have $[j^{-1}I_k,B]=0$, and its leading terms 
are $W$-invariant. Moreover, there exists a unique $\Delta _k\in D$ with 
the same leading terms [16]. By virtue of the uniqueness, from Corollary 2 
it follows that $\Delta _k=j^{-1}I_k$.

\section{Explicit Form of Radial Parts of Laplace Operators}
\setcounter{equation}{31}

In [19] we have obtained explicit formulas for certain integrals in 
${\tilde S}$. With the aid of the theorem presented in Sec. 4 it is easy to 
obtain formulas for the radial parts of Laplace operators: $\Delta _k=
\xi ^{-1}\,I_k\xi $.

Below we denote
\begin{eqnarray*}
 x(\xi )&=&\\
\mbox{I.} &&\qquad \xi ^{-1}\quad (X^0),\\
\mbox{II.} &&\qquad \mbox{sh}^{-1}\xi \quad (X^-), \\
\mbox{III.} &&\qquad \sin ^{-1}\xi \quad (X^+). \end{eqnarray*}
Let us note that the IM depends only on the function $x^2(\eta )=V(\eta )$ 
and its derivatives, whereas the constants $g$, $g_1$, and $g_2$ are related 
to the root multiplicities $m$, $m_1$, and $m_2$ by the formula (12).

{\bf a) Spaces with root systems of type $A_{n-1}$.} Here generators  are 
$\Delta _2=B$, $\Delta _3,\ldots ,\Delta _n$, $\Delta _k=\xi ^{-1}\,J_k\xi $,
\be J_n=\exp \left\{ -\,\frac{g^2}2\,\sum _{k\neq l}x^2(q_k-q_l)\,\frac
{\partial}{\partial p_k}\,\frac{\partial}{\partial p_l}\right\} p_1p_2
\cdots p_n. \ee

Let $Q=\sum _{j=1}^n q_j$. Then the other $J_k$ can be defined recursively\,
\footnote{\,\,In different form, the formulas for $J_k$ for systems of type 
$A_{n-1}$ have appeared for the first time in [13]. The proof that $J_k\in 
{\tilde S}_k$ can be found in [19].}
\be J_{k-1}=\frac{i}{k-n-1}\,[Q,J_k]. \ee
In lower dimensions, it is possible to present an explicit form of the 
operators $I_k$ with other leading terms. Let us denote by the brackets 
$\langle A_{kl}^r\rangle $ the trace of the matrix $A^r$, and let 
$p=\mbox{diag}\,(p_1,\ldots ,p_n)$. Then
\begin{eqnarray*}
I_3&=&\sum _{k=1}^n p_k^3+3g^2\sum _{k\neq l}x^2(q_k-q_l)\,p_l,\\
I_4 &=& \sum _{k=1}^n p_k^4+2g^2\sum _{k\neq l}x^2(q_k-q_l)(2p^2_l+p_kp_l)+
g^2\,\langle x^4(q_k-q_l)\rangle \\
&+& g^2\sum _{k\neq l}\left\{ 2\left[ x^2(q_k-q_l)\right] 'ip_l-
\left[ x^2(q_k-q_l)\right] ''\right\} ,\\
I_5 &=& \sum _{k=1}^n p_k^5+5g^2\sum _{k\neq l}x^2(q_k-q_l)(p^3_l+p_k^2p_l)+
5g^4\,\langle x^4(q_k-q_l)p_l\rangle \\
&+& 5g^2\sum _{k\neq l}\left\{ \left[ x^2(q_k-q_l)\right] 'ip_l^2-
\left[ x^2(q_k-q_l)\right] ''\,p_l\right\} . 
\end{eqnarray*}

There exist two infinite series and one exceptional symmetric space with 
a root system of type $A_{n-1}$\,\footnote{\,\,Here and below we shall not 
consider symmetric spaces with a complex group of motion for 
which $U(q)=0$ [3].}. 
Here we list the spaces of noncompact type (the notation has been adopted 
from [16]):

\bigskip
\begin{tabular}{|c|c|c|c|}
\hline
&&&\\
& $X^-$ & $g^2$ & Rank\\
&&&\\
\hline
&&&\\
$A\,I$ & $SL(n,R)/SO(n)$ & $-\,\frac14$ & $n-1$ \\
&&&\\
\hline
&&&\\
$A\,II$ & $SU^*(2n)/Sp(n)$ & 2 & $n-1$\\
&&&\\
\hline
&&&\\
$E\,IV$ & $E_6/F_4$ & 12 & 2 \\
&&&\\
\hline \end{tabular}

\bigskip
{\bf b) Spaces with root systems of type $B_n$, $C_n$, $BC_n$, and $D_n$}. 
Here the generators are $\Delta _2=B$, $\Delta _4,\ldots ,\Delta _{2n}$ (or 
$\Delta _n'$ for $D_n$). Only the operator $\Delta _4=\xi ^{-1}I_4\xi $ 
is known.
\begin{eqnarray*}
I_4 &=& 2\sum _{k=1}^n p_k^4+8g^2\sum _{k\neq l}\left[ x^2(q_k-q_l)+
x^2(q_k+q_l)\right] p_l^2 +8g_1^2\sum _{k\neq l} x^2(q_l)\,p_l^2 \\
&+& 8g_2^2\sum _{l=1}^n x^2(2q_l)p_l^2+4g^2\sum _{k\neq l}\left[ x^2(q_k-q_l)
-x^2(q_k+q_l)\right] p_kp_l\\
&+& 4g^2\sum _{k\neq l}\left[ x^2(q_k-q_l)+x^2(q_k+q_l)\right] 'ip_l+
g^4\,\langle \left[ x(q_k-q_l)\right] ^4\rangle \\
&-& 8g_1^2\sum _{k=1}^n\left[ x^2(q_k)\right] 'ip_k-16\,g^2_2\sum _{k=1}^n
\left[ x^2(2q_k)\right] 'ip_k\\
&-& 2g^2\sum _{k\neq l}\left[ x^2(q_k-q_l)+x^2(q_k+q_l)\right] ''-4g_1^2
\sum _{k=1}^n\left[ x^2(q_k)\right] ''\\
&-&16\,g_2^2 \sum _{k=1}^n\left[ x^2(2q_k)\right] ''. 
\end{eqnarray*}

Thus, for symmetric spaces of rank 2, with the exception of spaces with a 
group of motions $G_2$, we have  a complete system of generators. Here 
we list the nonisomorphic spaces of noncompact type [16], [18]:

\bigskip
\begin{tabular}{|c|c|c|c|c|c|}
\hline 
&&&&&\\
&&$X^-$&$g^2$ & $g_1^2$ & $g_2^2$\\
&&&&&\\
\hline
&&&&&\\
1& $BD\,I$ & $SO_0(n,2)/(SO(n)\otimes SO(2)),\,n>2$& $-\,\frac14$ & 
$\frac18\,(n-2)(n-4)$ & 0\\
&&&&&\\
\hline
&&&&&\\
2& $C\,II$ & $Sp(2,2)/(Sp(2)\otimes Sp(2))$ & $\frac34$ & 1&0\\
&&&&&\\
\hline
&&&&&\\
3& $A\,III$ & $SU(n,2)/(SU(n)\otimes U(2)),\,n>2$ & 0& $\frac12(n-2)^2$ & 
$-\frac12$\\
&&&&&\\
\hline
&&&&&\\
4& $C\,II$ & $Sp(n,2)/(Sp(n)\otimes Sp(2)),\,n>2$& 2& $2(n-1)(n-2)$ & 
$\frac32$\\
&&&&&\\
\hline
&&&&&\\
5& $D\,III$ & $SO^*(10)/U(5)$ &2& 2& $-\frac12$ \\
&&&&&\\
\hline
&&&&&\\
6& $E\,III$ & $E_6/(SO(10)\otimes SO(2))$ &6& 8& $-\frac12$ \\
&&&&&\\
\hline

\end{tabular}

\end{document}